\documentclass[fleqn,usenatbib]{mnras}

\usepackage{newtxtext,newtxmath}

\usepackage[T1]{fontenc}

\DeclareRobustCommand{\VAN}[3]{#2}
\let\VANthebibliography\thebibliography
\def\thebibliography{\DeclareRobustCommand{\VAN}[3]{##3}\VANthebibliography}

\usepackage{graphicx}	
\usepackage{amsmath}	

\newcommand{\deriv}[2]{\ensuremath{\frac{{\rm d}#1}{{\rm d} #2}}}
\newcommand{\rev}{}
\newcommand{\revmath}{}

\title[Radiative cooling approximations]{\rev Introducing two improved methods for approximating radiative cooling in hydrodynamical simulations of accretion discs}

\author[A. K. Young et al.]{
Alison K. Young,$^{1,2}$\thanks{E-mail: alison.young@ed.ac.uk (AKY)}
Maggie Celeste,$^{3}$
Richard A. Booth,$^{4}$
Ken Rice,$^{1,2}$
Adam Koval,$^{1,2}$
\newauthor
Ethan Carter,$^{5}$ and
Dimitris Stamatellos$^{5}$
\\
$^{1}$SUPA, Institute for Astronomy, University of Edinburgh, Blackford Hill, Edinburgh EH9 3HJ, UK\\
$^{2}$Centre for Exoplanet Science, University of Edinburgh, Edinburgh, EH9 3HJ, UK\\
$^{3}$Institute of Astronomy, University of Cambridge, Madingley Rd, Cambridge CB3 0HA \\
$^{4}$ School of Physics and Astronomy, Sir William Henry Bragg Building, Woodhouse Ln., University of Leeds, Leeds LS2 9JT, UK\\
$^{5}$ Jeremiah Horrocks Institute for Mathematics, Physics, and Astronomy, University of Central Lancashire, Preston PR1 2HE, UK \\
}

\date{Accepted XXX. Received YYY; in original form ZZZ}

\pubyear{2024}

\begin{document}
\label{firstpage}
\pagerange{\pageref{firstpage}--\pageref{lastpage}}
\maketitle

\begin{abstract}
The evolution of many astrophysical systems depends strongly on the balance between heating and cooling, in particular star formation in giant molecular clouds and the evolution of young protostellar systems. Protostellar discs are susceptible to the gravitational instability, which can play a key role in their evolution and in planet formation. The strength of the instability depends on the rate at which the system loses thermal energy. To study the evolution of these systems, we require radiative cooling approximations because full radiative transfer is generally too expensive to be coupled to hydrodynamical models. Here we present two new approximate methods for computing radiative cooling that make use of the polytropic cooling approximation. This approach invokes the assumption that each parcel of gas is located within a spherical pseudo-cloud which can then be used to approximate the optical depth. The first method combines the methods introduced by Stamatellos et al. and Lombardi et al. to overcome the limitations of each method at low and high optical depths respectively. The second, the ``Modified Lombardi'' method, is specifically tailored for self-gravitating discs. This modifies the scale height estimate from the method of Lombardi et al. using the analytical scale height for a self-gravitating disc. We show that the Modified Lombardi method provides an excellent approximation for the column density in a fragmenting disc, a regime in which the existing methods fail to recover the clumps and spiral structures. We therefore recommend this improved radiative cooling method for more realistic simulations of self-gravitating discs.
\end{abstract}




\section{Introduction}

Radiative transfer is an important aspect of many hydrodynamical simulations. Accurate treatment of heating and cooling is particularly important for systems undergoing gravitational instability or fragmentation, such as star-forming clouds and young, self-gravitating protoplanetary discs, in which the strength of the gravitational instability is a function of the thermal balance between shock heating and radiative cooling \citep{Gammie2001}. This determines whether the disc develops spiral arms, fragments or maintains a smooth, axisymmetric structure.

Early work using isothermal simulations suggested that a self-gravitating phase should be very short-lived, with the disc either fragmenting quickly or spiral instabilities acting to stabilize the disc's density profile \citep{laughlin1994}. However, simulations that implemented more detailed thermodynamics demonstrated that this was mostly a consequence of the isothermal assumption and that a self-gravitating phase may persist for many orbits \citep{lodatorice2004}. Now that we expect planet formation to begin very early when discs are likely to be self-gravitating \citep{nixon18}, we need robust models of self-gravitating discs to address key questions including how planetesimals grow in young discs. However, directly implementing radiative transfer in 3-D simulations is by no means trivial. A full treatment would require wavelength-dependent and potentially anisotropic absorption, scattering, and emission of photons; tracing of the photons' paths through the system; and variable opacities and emissivities for both dust and gas components. Additionally, explicit radiative transfer timesteps are prohibitively short relative to hydrodynamic timescales, particularly in optically thick locations, greatly increasing computation time \citep{Castor2004, Brandenburg2020}. 
These difficulties have prompted the development of a number of approximate radiative transfer methods, each depending on some simplifying assumptions and approximations. Such methods are naturally limited in accuracy and scope compared to a full radiative transfer scheme, but allow for the simultaneous simulation of both hydrodynamic and radiative processes.

One such method which has been used to study the effects of cooling in protoplanetary discs is the so-called $\beta$-cooling model, developed by \citet{Gammie2001}. This approach parameterises the local cooling time $t_c$ in terms of a $\beta$ parameter and the local Keplerian angular velocity $\omega_{\rm K}$, such that $t_c = \beta / \omega_{\rm K}$. The simplicity of this parameterisation and ease of implementation in hydrodynamic codes has made it a popular choice for studying the relationship between cooling and fragmentation in discs \citep[e.g.][]{Rice2003, Cossins2009, Meru2011, Boss2017}, but it does not paint a complete picture. There is no compelling argument for the $\beta$ parameter remaining constant throughout the extent of the disc \citep{mercer18}, and more realistic cooling methods yield significantly different results \citep{Johnson2003, Vorobyov2020}.

There are two main classes of more physically-motivated cooling methods for protoplanetary disc simulations: those that impose an external radiation field and calculate radiation transport through the disc; and those that make use of local parameters to estimate the thermal balance at a given location in the disc. Examples of the first approach include codes that make use of the flux-limited diffusion approximation \citep{Levermore1981}, such as \citet{ayliffe2011} and \citet{meru2010} with fixed prescribed temperatures at the edge of the model; or Monte-Carlo methods to transport photon ``packets'' throughout the disc, such as the coupled {\sc mcfost + phantom} model \citep{Pinte2006,price2018,nealon2020b}. For the coupled radiative transfer and hydrodynamics approach, a radiative equilibrium calculation is performed every few hydrodynamical timesteps (this is chosen to be as infrequent as possible to reduce the overheads), after which the temperatures are updated. Although the radiative transfer is more accurate than under the flux-limited diffusion approximation and can, for example, reproduce the effects of shadowing, the coupled method is not time-dependent due to the radiative equilibrium assumption. The outcome of this is that temperatures are likely to be under- or over-estimated in different regions.
The second approach -- estimating the cooling rate based on local properties -- is the focus of this paper. Existing methods are sufficient for modelling symmetrical systems, i.e. spherical cloud collapse, but are less accurate for modelling discs since they overestimate the optical depth in the disc. This then changes the cooling time, which in turn affects the evolution of self-gravitating discs. The use of Monte Carlo radiative transfer would allow for irregular structures but is very costly and currently imposes radiative equilibrium. As discussed above, it is crucial to have a robust cooling model for studying gravito-turbulent discs, therefore a more accurate approximation for the optical depth within discs and therefore the cooling rate is needed, without adding a large computational expense.

In section \ref{sec:existing methods} we first describe the methods that are currently used to approximate radiative cooling by assuming that each parcel of gas, or SPH particle, is embedded within a spherical gas cloud (``polytropic cooling'') and using this to estimate the optical depth. The limitations of these methods are detailed in \ref{sec:prev_limitations} and we further motivate the development of our new techniques. In section \ref{sec:combined} we introduce the first of our improved methods for approximating radiative cooling in discs, which combines the two. We present our second method in section \ref{sec:improved lombardi} which provides an alternative means of regularising the \citet{lombardi2015} method in the mid-plane by making use of disc-specific geometry.

\section{Existing polytropic cooling methods}
\label{sec:existing methods}

Much of what we present here will be based on simulations using Smoothed Particle Hydrodyamics (SPH) \citep[e.g.,][]{monaghan92}, but it is also applicable to other hydrodynamic formalisms. In SPH, each particle represents a parcel of gas and to estimate its radiative cooling rate we need to know its optical depth. Physically, this involves integrating the opacity and density along the path from the particle to the edge of the particle distribution (e.g., in the case of a disc, to the disc surface). The cooling approximations described below instead estimate a mean column density from the particle to the surface and combine this with mean opacity estimates. Two different approaches for estimating this mean column density are described, firstly using the gravitational potential \citep{stamatellos07} and secondly using the pressure gradient \citep{lombardi2015}.

\subsection{The Stamatellos Method: The gravitational potential approach}
\label{sec:stam_method}

\citet{stamatellos07} introduced a method for approximating the optical depth of an SPH particle from its gravitational potential. The gravitational potential energy is already calculated in SPH simulations that include the gas self-gravity, hence minimising additional computational expense. This approach treats each parcel of gas as if it were embedded within a spherically symmetric `pseudo-cloud’ with polytropic density and temperature profiles. The scale-length of the polytrope is determined by the gravitational potential at the location of the SPH particle. Because we don't know where this gas parcel lies within the pseudo-cloud, we must calculate a mass-weighted average of the pseudo-column density over all possible positions within it. The mean column density for particle $i$ is then given by

\begin{equation}
\label{eq:avcoldensity_stam}
    \Bar{\Sigma}_i = \zeta_n \left[ \frac{-\psi_i \rho_i}{4 \pi G}\right]^{1/2},
\end{equation}

where $\zeta_n$ is a constant that depends on the polytropic index (which is related to $n$; see \citealt{stamatellos07} for the full derivation), and $\psi_i$ and $\rho_i$ are the gravitational potential and density at the position of the particle respectively. Following \citet{stamatellos07} and \citet{forgan2009} we use $\zeta_2 = 0.368$. We now require a pseudo-mean mass opacity $\Bar{\kappa_i}(\rho,T)$, to estimate the mass weighted optical depth. The pseudo-mean opacity for particle $i$ is calculated by averaging over locations in the pseudo-cloud in the same manner as for the mean column density. This allows us to estimate the optical depth using

\begin{equation}
   \Bar{\tau_i} = \Bar{\Sigma}_i \Bar{\kappa_i}.
   \label{eq:coldepth}
\end{equation}

\subsection{The Lombardi Method: the pressure gradient approach}
\label{sec:lombardi_method}
A slightly different approach was introduced by \citet{lombardi2015} in which the pressure scale-height is used to estimate the optical depth rather than the gravitational potential. This method still assumes that each particle is embedded within a polytropic pseudo-cloud but by using the pressure gradient at the location of the particle, it performs better for non-spherical gas distributions.

The pressure scale-height, $H_{{\rm p}, i}$, of particle $i$ is given by:

\begin{equation}
    H_{{\rm p},i} \equiv P_i/\left| \nabla P_i \right|,
\end{equation}
where $P_i$ is the pressure at the location of particle $i$, and $\nabla P_i$ is the pressure gradient.

The hydrodynamic acceleration excluding the contributions from artificial (and real) viscosity and gravity is

\begin{equation}
    \boldsymbol{a}_{{\mathrm h},i} = - \frac{\nabla P_i}{\rho_i}.
\end{equation}

In smoothed particle hydrodynamics, the hydrodynamic acceleration $\boldsymbol{a}_{{\mathrm h},i}$ is already computed during the force calculation. For particle $i$, this is calculated via the following sum over $j$ neighbour particles:

\begin{equation}
    \boldsymbol{a}_{{\mathrm h},i} \equiv \frac{d\boldsymbol{v}_i}{dt} = -\sum_j \left [  \frac{m_j P_i}{\rho_i^2 \Omega_i} \nabla W_i + \frac{m_j P_j}{\rho_j^2 \Omega_j} \nabla W_j  \right ]   ,
\end{equation}

in which $m_j$ is the particle mass, $\rho_i$ ($\rho_j)$ is the gas density evaluated at particle $i$ ($j$), and the gradient of the smoothing kernel is $\nabla W$. $\Omega $ is related to the gradient of the smoothing length $h_i$

\begin{equation}
    \label{eq:omega}
    \Omega_i \equiv 1 - \frac{\partial h_i}{\partial \rho_i} \sum_j m_j \frac{\partial W_{ij}(h_i)}{\partial h_i}.
\end{equation}

We can then obtain the pressure scale-height via

\begin{equation}
      H_{{\rm p},i} = \frac{P_i}{\rho_i \left| \boldsymbol{a}_{{\mathrm h},i} \right|}.
\end{equation}

Calculating the mass-weighted pseudo-mean column density in the same manner as \citet{stamatellos07}, \citet{lombardi2015} obtain

\begin{equation}
\label{eq:coldens_lom}
    \Bar{\Sigma}_i = \zeta ' \rho_i H_{{\rm P}, i} = \frac{\zeta ' P_i }{\left | \boldsymbol{a}_{{\rm h},i} \right |},
\end{equation}

with $\zeta ' = 1.014 $.  The optical depth can then be estimated using Equation \ref{eq:coldepth}.

\subsection{Equation of state}

We employ an identical approach to \citet{stamatellos07} in which values of the pseudo-mean opacity $\Bar{\kappa_i}$, Planck mean opacity $\kappa_i$, mean molecular weight $\mu_i$, and specific internal energy $u_i$ are pre-computed for the required density and temperature ranges and stored in a look-up table to reduce computational expense. These were calculated following \citet{bell1994}. The Rosseland-mean and Planck-mean opacities are assumed to be the same and to depend only on density and temperature and are given by:
\begin{equation}
    \label{eq:meanopacs}
    \kappa_{\rm R} (\rho,T) = \kappa_{\rm P}(\rho,T) = \kappa_0 \rho^a T^b,
\end{equation}
where the constants $\kappa_0$, $a$ and $b$ are chosen for each main physical process contributing to the total opacity in that temperature and density range (see \citealt{bell1994,stamatellos07} for further details). The tabulated values are then interpolated during the live hydrodynamical computation and the ideal gas approximation is used to obtain the gas pressure.

\subsection{Updating the energy}

Both methods give us a mass-weighted column density and tabulated values of the pseudo-mean opacity. These quantities are now used to calculate the radiative heating rate of gas parcel represented by particle $i$:

\begin{equation}
    \label{eq:coolingrate}
    \frac{du_i}{dt}\Big| _{\rm rad} = \frac{4 \sigma \left(T_{0}^4(\boldsymbol{r}_i) - T_i^4 \right)}{\Bar{\Sigma}^2\Bar{\kappa_i}(\rho_i,T_i) + \kappa_i^{-1}(\rho_i,T_i)},
\end{equation}
where $T_0$ is a minimum temperature determined by background radiation and $\kappa_i^{-1}(\rho_i,T_i)$ is the Planck mean opacity, which is tabulated along with the mass-weighted opacity $\Bar{\kappa_i}(\rho_i,T_i)$. External heating from nearby stars can be included by calculating $T_0$ for each gas particle to prevent the gas cooling radiatively below this minimum temperature.\footnote{\citet{lombardi2015} introduce a slightly different method for calculating the radiative heating rate but we opt to use the original approach of \citet{stamatellos07} for both methods.}

The equilibrium temperature of particle $i$ is found by assuming that the radiative cooling and the compressive and viscous heating are in balance:

\begin{equation}
  \frac{du_i}{dt}\Big| _{\rm hydro}  + \frac{du_i}{dt}\Big| _{\rm rad}  = 0.
\end{equation}
Substituting the equilibrium temperature $T_{{\rm eq},i}$ for $T_i$ in Equation~\ref{eq:coolingrate} gives:

\begin{equation}
    T_{\rm{eq},i}^4 =  \frac{1}{4\sigma}\left(\Bar{\Sigma}^2\Bar{\kappa_i}(\rho_i,T_i) + \kappa_i^{-1}(\rho_i,T_i)\right)\frac{du_i}{dt}\Big| _{\rm hydro} + T_{0}^4(\boldsymbol{r}_i).
\end{equation}

In the case that $\revmath T_{\rm{eq},i}^4 \leq T_{0}^4(\boldsymbol{r}_i)$, we set $T_{\rm{eq},i}^4 = T_{0}^4(\boldsymbol{r}_i)$. The equilibrium energy is then $u_{{\rm eq},i} = u(T_{{\rm eq,}i},\rho_i) $. The energy is updated by considering the approach towards the equilibrium temperature with thermal timescale
\begin{equation}
    t_{{\rm therm},i} = \left(u_{\rm{eq},i} - u_i \right) \left [\frac{du_i}{dt}\Big| _{\rm hydro} + \frac{du_i}{dt}\Big| _{\rm rad} \right ]^{-1}.
\end{equation}
%
For implementation within a typical SPH code, the cooling rate is then given by
\begin{multline}
\label{eq:coolingrate_phantom}
    \frac{du_i}{dt}\Big| _{\rm cool} = \frac{1}{\delta t} \left[ u_i \exp \left(
    \frac{-\delta t}{t_{\rm{therm},i}}\right) 
    + u_{\rm{eq},i} \left ( 1 - \exp\left(\frac{-\delta t}{t_{\rm{therm},i}}\right) \right ) \right] \\
    + \frac{du_i}{dt}\Big| _{\rm hydro} .
\end{multline}
The energy equation is then updated via
\begin{equation}
    u_i (t+\delta t) = u_i(t) + \delta t  \frac{du_i}{dt}\Big| _{\rm cool}.
\end{equation}

\subsection{Limitations of these methods}
\label{sec:prev_limitations}

 In both cases, given the estimate is based only on local parameters already used in hydrodynamic calculations in SPH, there is no need for communication of information with other fluid elements, and there is therefore a minor computational cost when compared with methods that utilise a radiation field or photon packets. However, any local approximation of optical depth will necessarily require assumptions about the system's global geometry. In the case where one wishes to study systems with multiple geometries co-occurring -- for example, discs with spherical clumps and spiral waves, or with low-density envelopes or streamer tails -- accurately estimating the cooling rate in all regions of the disc with just one method can be problematic.

The \citet{stamatellos07} method significantly over-estimates the optical depth in lower density regions of the disc but performs better towards the disc mid-plane and in spherical clumps. The \citet{lombardi2015} method is more accurate overall in disc simulations \citep{mercer18}, but breaks down in regions where the pressure gradient approaches zero, such as in the disc mid-plane or the center of clumps. By combining the two methods, one can take advantage of the strengths of each approach to achieve more accurate cooling throughout the whole of the disc. It is also possible to improve on the \citet{lombardi2015} method in the regions where it fails by using another means of estimating the optical depth that makes use of a disc-specific density distribution assumption, as there is a simple relationship between mid-plane density, scale-height and column density in discs.

\section{Method}
\label{sec:method}

In this section we detail two new methods for estimating the optical depth of a particle. First, in \ref{sec:combined} we combine the estimates from the methods of \citet{stamatellos07} and \citet{lombardi2015}, and second in \ref{sec:improved lombardi} we modify the method of \citet{lombardi2015} specifically tailored for modelling self-gravitating discs.

\subsection{The Combined Method: combining estimates of optical depth}
\label{sec:combined}

\begin{figure}
    \centering
    \includegraphics[width=\columnwidth]{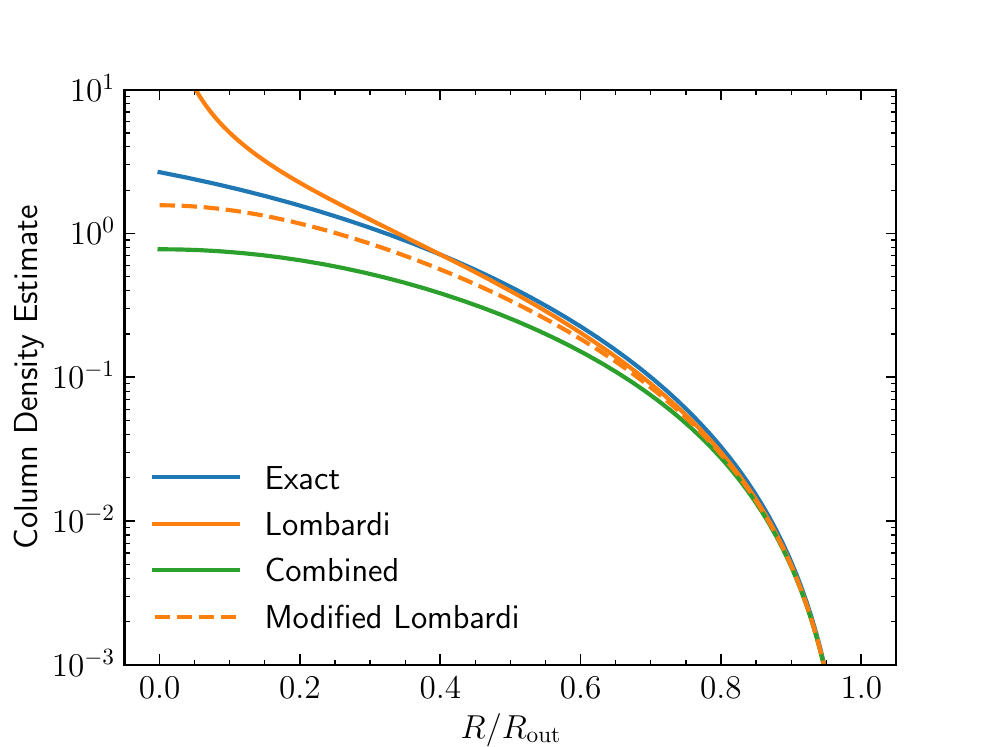}
    \caption{The column density from $R$ to $R_{\rm out}$ for an $n=3/2$ polytrope (blue line, "Exact") compared with the estimate from \citet{lombardi2015}. Also shown are the estimate generated by the combined method (see \autoref{sec:combined}) and a modification of the Lombardi method for discs (see \autoref{sec:improved lombardi}).}
    \label{fig:polytrope}
\end{figure}

\begin{figure}
    \centering
    \includegraphics[width=\columnwidth]{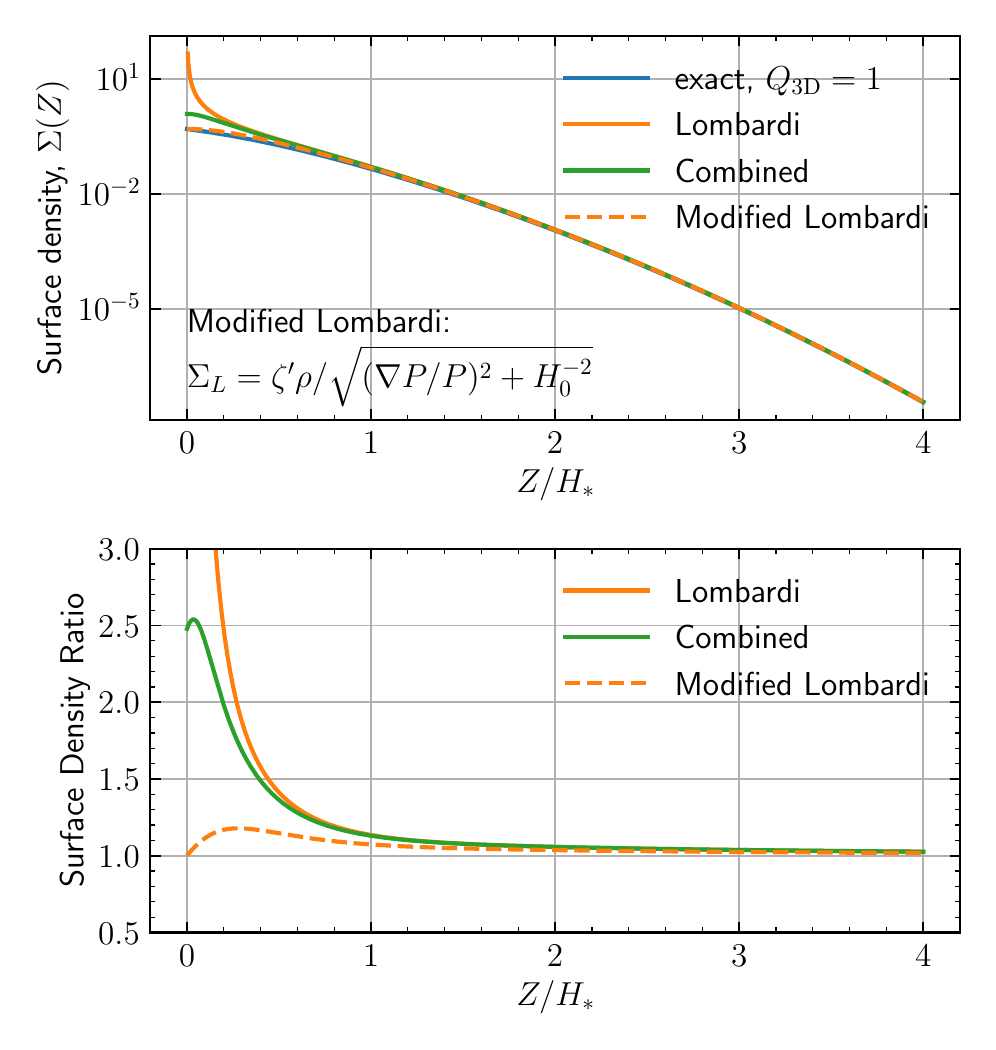}
    \caption{Top: The column density from $z$ to infinity for the analytic self-gravitating disc model. Also shown are different estimates based on the Lombardi method and Combined method, as well as a Modified Lombardi method applicable to discs (see \autoref{sec:improved lombardi}). Bottom: The ratio of the estimated column density to the exact column density.}
    \label{fig:disc_1d}
\end{figure}

While \citet{lombardi2015} and \citet{mercer18} showed that the pressure scale-height and density can be used to provide a good estimate of the column density (\autoref{eq:coldens_lom}), this estimate breaks down in regions where the pressure gradient is close to zero, such as at the centre of a fragment or close to a disc mid-plane. Since a large part of the mass can be found under such conditions, the average cooling rate can be affected and we are motivated to find a way to improve the estimate under these conditions.

Here we explore replacing the Lombardi method with a surface density estimated by the Stamatellos method in regions where $\nabla P \rightarrow 0$ as the Stamatellos method remains well defined in such regions.

First, we define a scale-height based on the Stamatellos method via
\begin{equation}
    H_{\rm S} = \frac{\zeta_n}{\zeta'} \left[\frac{-\psi}{4\upi G\rho}\right]^{1/2},
\end{equation}
which we combine with the Lombardi estimate $\revmath{H_{\rm P}}$ in inverse quadrature, i.e. 
\begin{equation}
    H_{\rm C} = \left(H_{\rm P}^{-2} + H_{\rm S}^{-2}\right)^{-1/2}. \label{eq:combined}
\end{equation}
This selects the smaller of the two estimates, chosen because the Stamatellos method is known to overestimate the surface density in disc geometries and therefore generally we will only want to use this estimate when the Lombardi method is also failing, producing overestimates itself.

As a first test of this Combined method we consider a polytrope as an approximation to the structure of a self-gravitating fragment, and show $\Sigma(z)$ for an $n=3/2$ polytrope and also the different column density estimates in \autoref{fig:polytrope}. As previously stated, the Lombardi method works well away from the centre of the polytrope but overestimates the column density once the density and pressure gradients tend to zero close to the centre of the polytrope.  The Combined method offers an improvement over the Lombardi method alone near the core, but it does lead to a moderate underestimate of the column. This follows because the Stamatellos method uses an estimate of the column density that is averaged over the entire polytrope and therefore must underestimate the column at the centre, where the column is highest. 

Next, we consider how the Combined method works in a disc geometry. To this end, we consider a thin, self-gravitating disc. High-resolution simulations show that self-gravitating discs have a temperature structure that is typically close to vertically isothermal \citep{Shi2014, Booth2019}, so we choose a vertically-isothermal structure. Then vertical hydrostatic equilibrium implies 
\begin{multline}
    \frac{1}{P}\deriv{P}{z} = \frac{1}{\rho}\deriv{\rho}{z} = - \frac{\omega_{\rm K}^2 z + 4\upi G \rho_0 \tilde{\Sigma}}{c_{\rm s}^2} \\
    = -\frac{1}{H_*^2}\left(z + \frac{\tilde{\Sigma}}{Q_{\rm 3D}}\right) =  -H_{ P}^{-1}, \label{eq:disc_1d_H}
    \end{multline}
where we have assumed that the disc's self-gravity can be calculated using a slab approximation, as in \citet{wilkins2012} for example. Here $H_*=c_{\rm s}/\omega_{\rm K}$, where $c_{\rm s}$ is the isothermal sound speed, $\omega_{\rm K}$ is the Keplerian angular frequency, and $\tilde{\Sigma} = \int_0^z \rho(z)/\rho(0) {\rm d} z$. For the last equality we have introduced $Q_{\rm 3D} = \omega_{\rm K}^2/4\upi G\rho(0)$, which is the 3D definition of the Toomre Q parameter \citep[e.g.][]{mamatsashvili10,lin_kratter2016}. 

To compute the potential we need to extend this 1D model to a global disc because the potential of an infinitely extended self-gravitating slab is undefined. We choose a disc that has a constant $Q_{\rm 3D}$ with radius so the vertical structure is given by the same solution to \autoref{eq:disc_1d_H} everywhere. The definition of $Q_{\rm 3D}$ then enforces the mid-plane density scaling as $\rho_0 \propto R^{-3}$. Specifying the scale-height, $H_*$,  fixes the surface density and the disc model is complete. Here we choose $H_*/R = 0.1$. We then compute the potential by direct integration of Green's function for Poisson's equation in 3D. We assume that the disc extends from 5~au to 100~au for consistency with the discs used in the SPH simulations. We note that while the final potential is not perfectly consistent with the disc structure, the difference is small when the disc is thin \citep[e.g.][]{bertin99,wilkins2012}

In \autoref{fig:disc_1d} we compare the surface density for the analytic disc model to the Lombardi estimate using the exact $H_{\rm P}$ from \autoref{eq:disc_1d_H}. The estimate works well at large $z$ but, as expected, it overestimates the surface density at $z=0$. To regularize the Lombardi method we evaluate the potential assuming a point 50~au from the star. Here, we see that the Combined method improves on the Lombardi estimate, as expected. In this case, the Combined method moderately overestimates the column at the mid-plane, which is expected since the Stamatellos method has been shown previously to overestimate the column in disc geometries \citep{wilkins2012, lombardi2015, mercer18}. 

These two simple tests demonstrate that combining the Lombardi and Stamatellos methods should offer a significant improvement over using either method alone. In section \ref{sec:tests} we explore this for more realistic scenarios that arise in SPH simulations.

\subsection{The Modified Lombardi method: An improved Lombardi method for disc simulations}
\label{sec:improved lombardi}

The Stamatellos method is not the only viable option for regularizing the Lombardi method.  Since the Lombardi method only fails in regions where $\nabla P \approx 0$, such as the mid-plane of discs or the centre of polytropes, we only need to replace the estimate at such locations. Here we explore an alternative that makes use of the fact that there is a simple relationship between the mid-plane density, scale-height and column density in a disc geometry.

\begin{figure}
    \centering
    \includegraphics[width=\columnwidth]{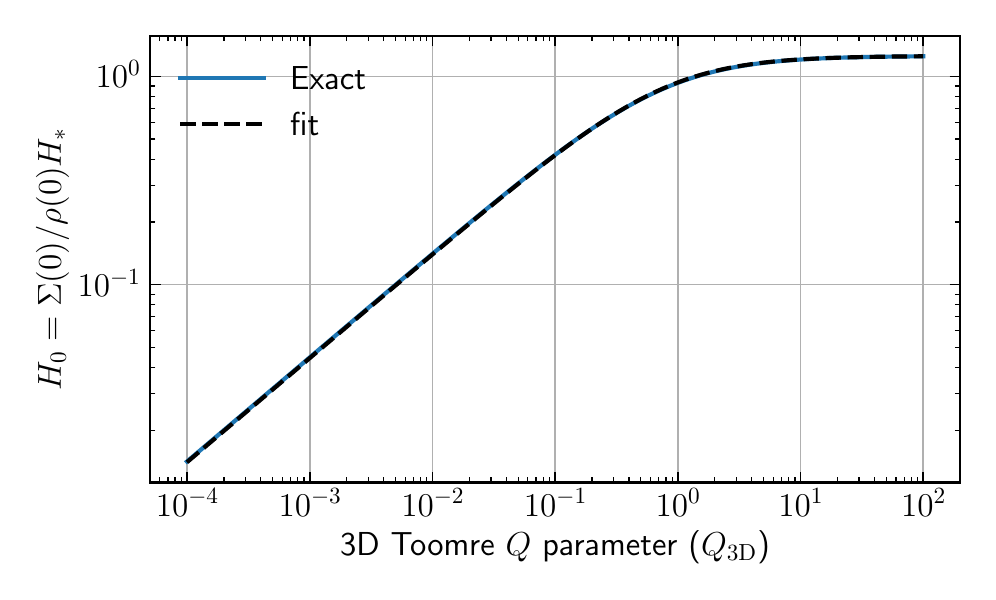}
    \caption{A comparison between the half thickness of a self-gravitating disc for different $Q_{\rm 3D}$ and the fit formula given in \autoref{eq:disc_H_approx}.}
    \label{fig:H_mid}
\end{figure}

For a non-self-gravitating disc, the (half) column density and mid-plane density are related via $\Sigma/2 =  \rho_0 \sqrt{\upi/2} H_*$, and so $H_0 = \sqrt{\upi/2} H_*$ could be used in the place of $H_{\rm S}$. Away from the mid-plane $H_{\rm P} < \sqrt{\upi/2} H_*$, so the Lombardi method would take over. However, using $\sqrt{\upi/2} H_*$ overestimates the column density for self-gravitating discs or in fragments because self-gravity further compresses the gas, reducing the scale-height. Fortunately, it is possible to estimate $H_0$ for a disc with any $Q_{\rm 3D}$. For $Q_{\rm 3D} \gg 1$, $H_0 = \sqrt{\upi/2}H_*$ suffices. In the regime $Q_{\rm 3D} \ll 1$ the second term in \autoref{eq:disc_1d_H} dominates and the equation can be made dimensionless by substituting $z$ with $z'\lambda$ where $\lambda = H_*\sqrt{Q_{\rm 3D}}$. We therefore expect $H_0/H_*$ to scale with $\sqrt{Q_{\rm 3D}}$ when $Q_{\rm 3D}$ is small, and indeed find the fitting formula,
\begin{equation}
\frac{H_0}{H_*} = \frac{t}{\sqrt{1 + 1/(tQ_{\rm 3D})}}, \label{eq:disc_H_approx}
\end{equation}
describes {\rev $\Sigma(0)/\rho(0)$} well, where $t = \sqrt{\upi/2}$ (\autoref{fig:H_mid}) \footnote{\rev In the SPH model, we implement $\omega_{{\rm K},i}=\sqrt{G M_*/|\boldsymbol{r}_{K,i}|^3}$ where $\boldsymbol{r}_{K,i}= \boldsymbol{r}_i - \boldsymbol{r}_*$ is the separation between the gas particle $i$ and the sink particle (star). If there is no sink particle, $\omega_{{\rm K},i}=0$. This approach would need to be adapted for simulations with multiple stars, for example by adding $\omega_{{\rm K},i}$ for each sink particle in quadrature.}.

An advantage of \autoref{eq:disc_H_approx} is that it can be evaluated using quantities local to the disc mid-plane only (i.e., $\rho$, $c_{\rm s}$, and $\omega_{\rm K}$). Further, while \autoref{eq:disc_H_approx} is only strictly meaningful at the mid-plane it can be evaluated anywhere as it only depends on local quantities. Since $H_0$ increases away from the mid-plane (because $\rho$ decreases) 
and $H_{\rm P}$ decreases, replacing $H_{\rm S}$ with \autoref{eq:disc_H_approx} in \autoref{eq:combined} results in a method that produces similar estimates to the Lombardi method away from the mid-plane, which is generally quite accurate. Indeed, \autoref{fig:disc_1d} shows that this `Modified Lombardi' method outperforms the normal Lombardi and Combined Lombardi + Stamatellos methods in discs. For the method to be robust, however, we need it to work well when the geometry deviates significantly from a disc.

Next, we apply this Modified Lombardi method to a polytrope as a model for a fragment in a self-gravitating disc. Under such conditions, we expect $Q_{\rm 3D} \ll 1$, such that $H_0 \approx 2 c_{\rm s} / \sqrt{4\upi G\rho_0}$. We apply this estimate of $H_0$ to the polytrope in \autoref{fig:polytrope}, which shows that \autoref{eq:disc_H_approx} provides an estimate of $\Sigma(z)$ that is accurate to within a factor of 2, in a spherical geometry. Since the size of a polytrope scales with $c_{\rm s} \sqrt{(n+1)/4\upi G \rho(0)}$ \citep[e.g.][]{Clarke2014}, the accuracy of this estimate is independent of $Q_{\rm 3D}$ as long as $Q_{\rm 3D} \ll 1$. Since the modified Lombardi method works well for disc geometries in general and for polytropic clumps when $Q_{\rm 3D} \ll 1$, we expect it to be widely applicable to the structures present in self-gravitating discs {\rev and it also works well for spherical geometries}. Finally, it should be noted that in deriving \autoref{eq:disc_H_approx} we have assumed hydrostatic equilibrium. As a result, the Modified Lombardi method may not offer a significant improvement away from hydrostatic equilibrium.


\section{Tests}
\label{sec:tests}

We now present tests using the SPH code {\sc phantom} \citep{price2018}. For comparisons with previous work, we first simulate the collapse of a uniform density sphere and then consider both low- and high-mass discs.  For the collapse of the uniform density sphere, we only consider the Combined method, but for the low- and high-mass protoplanetary discs we consider both the Combined method and the Modified Lombardi method. The key factor that determines how accurately the cooling rate is estimated and the difference between the methods is how well the column density at each position is estimated. The column density estimates are therefore the focus of this section. 

\subsection{Collapse of a uniform density sphere}

The gravitational collapse of a uniform density sphere represents the collapse of a protostellar cloud core and forms a useful benchmark (see \citealt{masunaga00} for the first radiation hydrodynamic simulation of such a collapse).
For this test, we use $2 \times 10^6$ SPH particles to simulate a cloud core of mass $1.5$~M$_\odot$, with an initial radius of $10^4$~au, and an initially uniform temperature of $5$~K.

\begin{figure}
    \centering
    \includegraphics[width=0.9\linewidth]{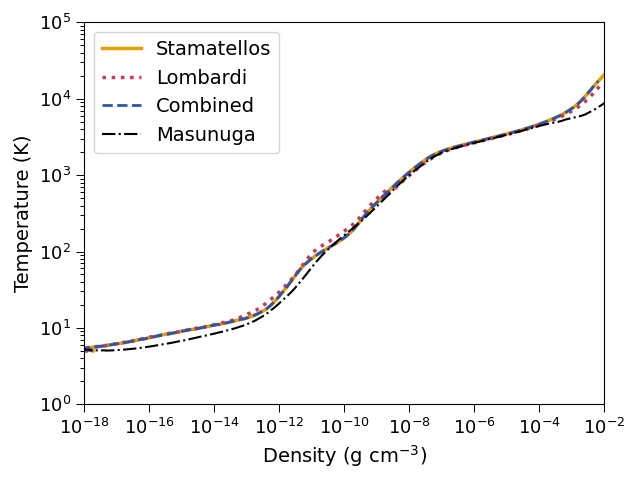}
    \caption{The evolution of the maximum density and temperature during the collapse of a uniform density sphere with three radiative cooling methods: Stamatellos method (yellow line), Lombardi method (red dotted line), and the Combined method (blue dashed line). The results from the simulation of \citet{masunaga00} are plotted (black dashed-dot line) for comparison.
    The initial mass and radius are $1.5$~M$_\odot$ and $10^4$~au respectively, and the temperature is initially 5~K throughout. The evolution under the Combined method lies almost exactly on top of the results obtained using the Stamatellos method.}
    \label{fig:cloud_collapse}
\end{figure}

Fig. \ref{fig:cloud_collapse} shows the evolution of the maximum density and temperature of the cloud core using the Stamatellos method, the Lombardi method, and the Combined method. 
For the Stamatellos and Combined methods, this is simply the density and temperature of the SPH particle with the highest density. For the Lombardi method there is considerable scatter so we average the 200 SPH particles with the highest density. 
The evolution of models with the Stamatellos, Lombardi and Combined methods are very similar. Moreover, Figure \ref{fig:cloud_collapse} shows that these methods produce results that are close to what is expected (black dash-dot line from \citealt{masunaga00}). As mentioned in \citet{stamatellos07}, the differences with respect to \citet{masunaga00} are likely due to the different opacities implemented.

The small differences in the evolution between the models are due to the methods used to estimate the column density. Therefore, we now examine the column density estimates in each case. For this analysis we used a single simulation (using the Combined cooling method) and determined the column density estimate for all three cooling methods at different times in this simulation.

\begin{figure}
    \centering
\includegraphics[width=\columnwidth]{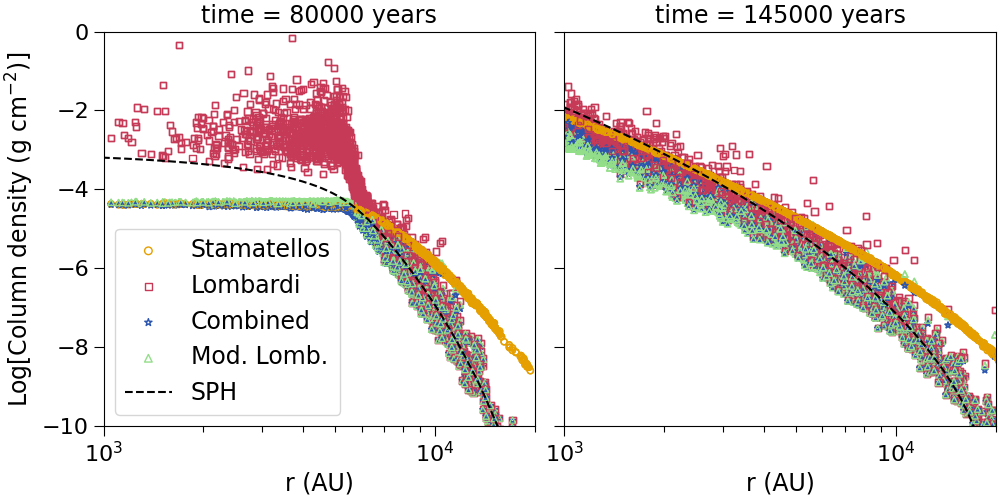}
    \caption{Estimates of the column density for a collapsing  sphere with an initially uniform density at $t = 80000$~years (left-hand panel) and at $t=145000$~years (right-hand panel). Estimates are calculated using the Stamatellos method (yellow open circles), the Lombardi method (red open squares), the Combined method (blue stars), and the Modified Lombardi method (green triangles). The dashed line is the ``true'' column density profile, determined by integrating the density in the SPH simulation from the centre of the cloud to the surface.}
    \label{fig:coldens_collapse}
\end{figure}

Figure~\ref{fig:coldens_collapse} shows the resulting column density estimates as a function of radius for two snapshots during the collapse. We compare these estimates with the ``true'' column density (black dashed lines), calculated by integrating the density outwards from the specified cloud radius to the surface. 
Near the cloud core centre, the Lombardi method initially over-estimates the column density and Fig.~\ref{fig:coldens_collapse} displays a lot of scatter. This is because the pressure gradient will initially be close to zero near the core centre and so the Lombardi method will not produce reliable estimates of the column density. It is because of this scatter that it is necessary to average the 200 particles with the highest density when producing the results for the Lombardi shown in Fig.~\ref{fig:cloud_collapse}. 

On the other hand, near the cloud centre, the Stamatellos method, the Combined method (which tends to the Stamatellos method in these regions) {\rev and the Modified Lombardi method} initially under-estimate the column density, but avoid the large scatter seen in the Lombardi method.  However, at larger radii, the Stamatellos method clearly over-estimates the column density, while the {\rev Lombardi and Modified Lombardi methods produce values much closer to the real column density}.  In these regions, the Combined method tends to the Lombardi method and produces a reliable estimate of the column density.

The right-hand panel of Figure~\ref{fig:coldens_collapse} shows that as the cloud collapse progresses, the Combined method tends to the Lombardi method throughout most of the cloud core and both return reasonable estimates of the column density.{\rev The Modified Lombardi method also provides a reasonable estimate of the column density, despite being designed for a disc geometry.}

\subsection{Evolution of a protoplanetary disc}

\subsubsection{Low-mass disc}
\label{sec:low-mass}
We first consider a low-mass disc, similar to that presented in \citet{mercer18}. The disc has an initial surface density profile of $\Sigma \propto r^{-1}$, extending from $r = 5$~au to $r = 100$~au, and an initial sound speed profile of $c_{\rm s} \propto r^{-0.25}$. The total disc mass is $M_{\rm disc} = 0.01$~M${_\odot}$, and the aspect ratio at $r= 5$~au is initially $H/r = c_{\rm s}/\Omega = 0.05$.  The disc gas is represented by $2 \times 10^6$ SPH particles and the central star has a mass of $M_* = 0.8$~M$_\odot$ and is represented by a sink particle. The artificial viscosity parameters are $\alpha_{\rm SPH}$ = 1.0 and $\beta_{\rm SPH} = 2.0 $.

\begin{figure}
    \centering
    \includegraphics[width=0.9\linewidth]{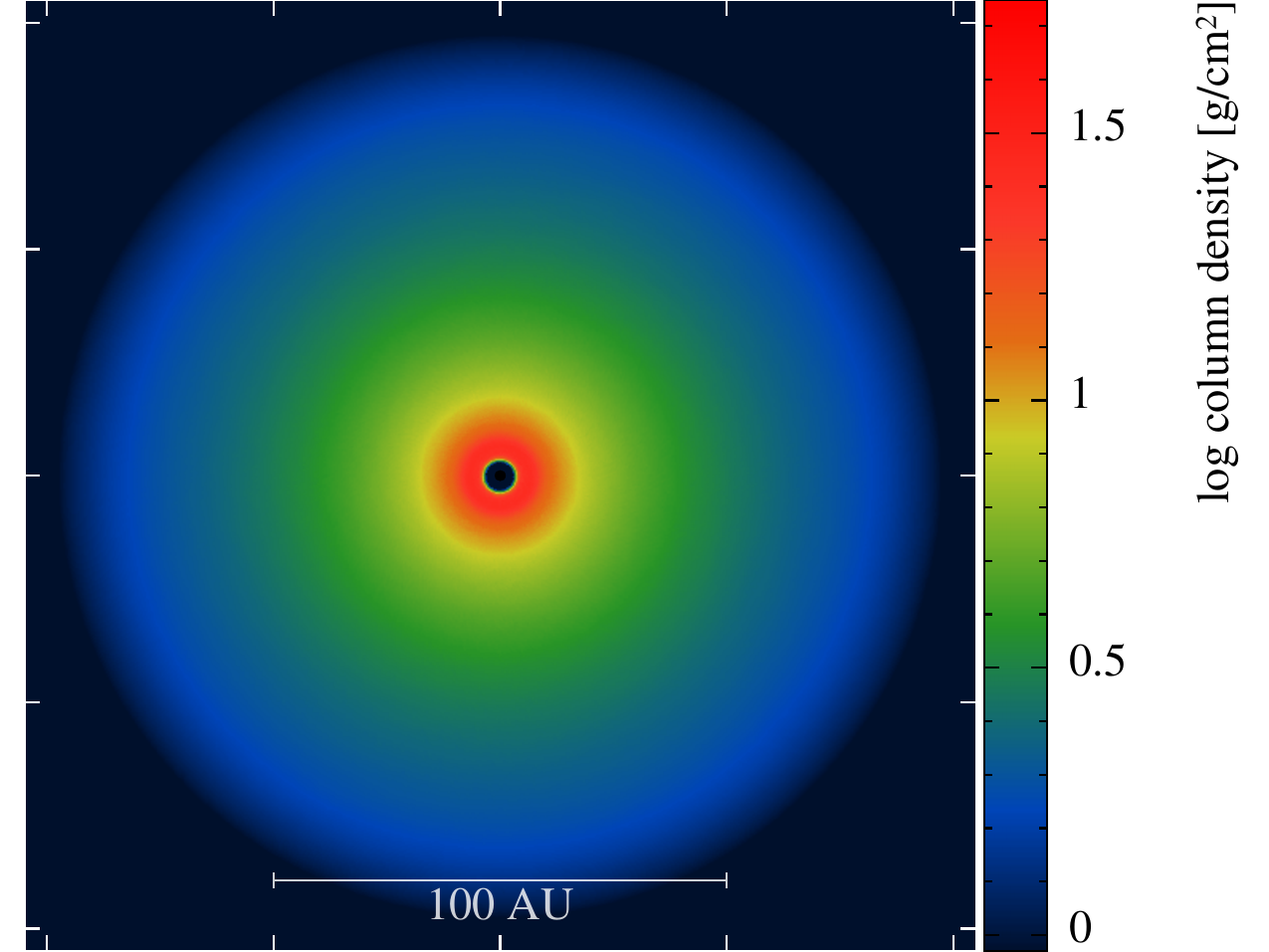}
    \caption{Snapshot of the disc column density at $t = 1200$ years for the $M_{\rm disc} = 0.01$~M$_\odot$ (low-mass) disc discussed in Section \ref{sec:low-mass}.}
    \label{fig:low-mass-disc-coldens}
\end{figure}

\begin{figure}
    \centering
    \includegraphics[width=0.9\linewidth]{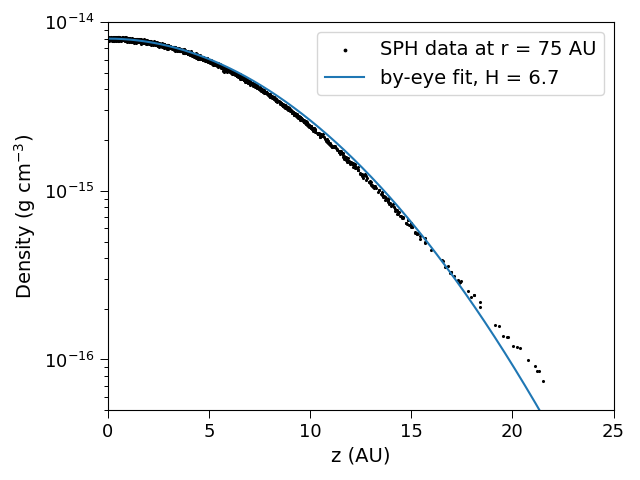}
    \caption{By-eye fit to the vertical profile, at $r = 75$~au, of the SPH simulation of a 0.01 M$_\odot$ (low-mass) disc around a 0.8 M$_\odot$ central star.}
    \label{fig:vertprofile}
\end{figure}

We run the simulation for $t = 1200$~years to allow the disc to settle and specify a minimum disc temperature of $T_0 = 25$~K. The column density snapshot of the disc at $t = 1200$~years (Fig.~\ref{fig:low-mass-disc-coldens}) shows the disc has a smooth structure. We then consider a thin azimuthal ring within the disc at $ r = 75$~au.  Fig.~\ref{fig:vertprofile} shows the vertical density structure at $r = 75$~au with a by-eye fit to the particle plot. Fitting with the expected profile of $\rho = \rho_o \exp(-z^2/2H^2)$ demonstrates that the scale-height at $r = 75$~au is about $H = 6.7$~au. 

\begin{figure}
    \centering
    \includegraphics[width=0.9\linewidth]{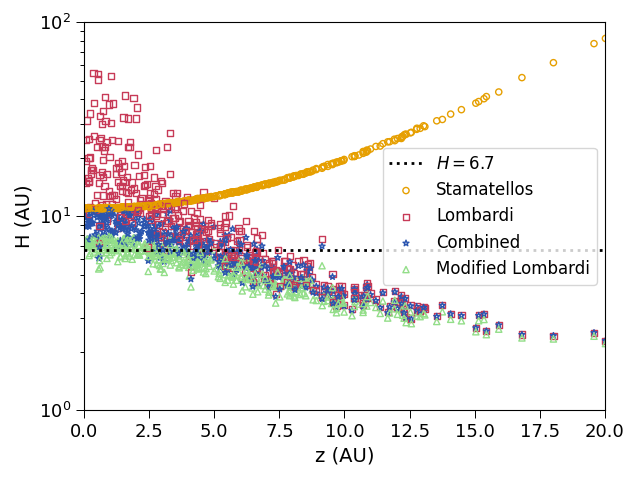}
    \caption{Estimates for the disc scale-height at $r = 75$ AU of the low-mass disc calculated with the Stamatellos method (yellow circles), the Lombardi method (red open squares), the Combined method (blue stars) and the Modified Lombardi method (green triangles).The dotted line is $H = 6.7$~AU, the ``true'' scale-height obtained from the fit shown in Figure~\ref{fig:vertprofile}.}
    \label{fig:H_estimates}
\end{figure}

We now compare this value obtained directly from the simulation to estimates obtained using each of the Stamatellos, Lombardi, Combined, and Modified Lombardi   methods. Estimates of the scale-height, $H$, for a randomly selected sample of SPH particles are presented in Fig.~\ref{fig:H_estimates}, along with the value obtained directly from the simulation ($H=6.7$~AU, dotted line, c.f., Fig.~\ref{fig:vertprofile}). Near the midplane, the Stamatellos method (open yellow circles) produces a reasonable estimate for the scale-height, but at higher altitudes is clearly far too large.  The Lombardi method is, again, unreliable near the midplane, since this is a region where the pressure gradient is tending towards zero, but appears to produce more reasonable estimates at higher altitudes.  Combining the two methods (Fig.~\ref{fig:H_estimates}, blue stars) then produces estimates that tend to the Stamatellos method near the midplane, and to the Lombardi method at higher altitudes. The Modified Lombardi method is an almost exact match to the expected scale-height in the midplane, and then follows the Lombardi method at higher altitudes.  

\begin{figure}
    \centering
    \includegraphics[width=0.9\linewidth]{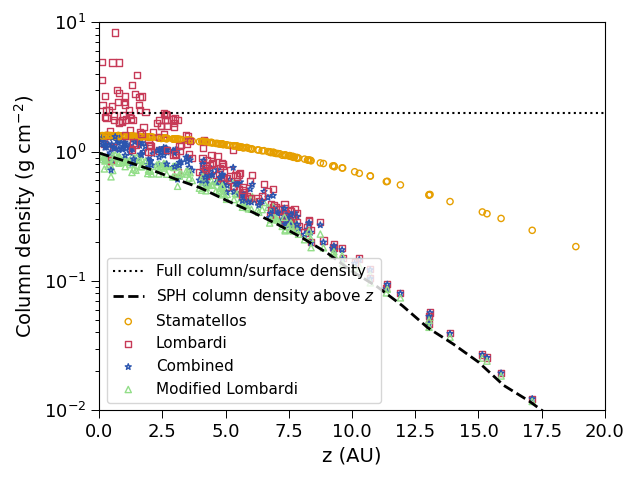}
    \caption{Estimates for the column density against $z$ at $r = 75$~AU of the $0.01$ M$_\odot$ (low-mass) disc using the Stamatellos method (yellow open circles), the Lombardi method (red open squares), the Combined method (blue stars) and the Modified Lombardi method (open green triangles). The dashed line shows the column density determined directly from the simulation by integrating the density from $z$ to the nearest disc surface. The total column density at $r = 75$~AU (equivalent to the surface density) is indicated by the dotted line and will be twice the column density from the midplane to the nearest surface. We are looking to obtain estimates as close to the dashed line as possible.}
    \label{fig:column_dens_estimates}
\end{figure}

Fig.~\ref{fig:column_dens_estimates} shows the column density estimates found using equations \ref{eq:avcoldensity_stam} and \ref{eq:coldens_lom} at $r = 75$~AU as a function of the height above the midplane, $z$, for the four methods together with the full surface density from the SPH simulation (dotted line) and column densities determined directly from the SPH simulation (dashed line). Again, for the Lombardi, Stamatellos, Combined, and Modified Lombardi methods, we plot a random sample of the estimates. The full SPH surface density (dotted line) is determined by integrating the SPH density at $r = 75$~AU over the full disc column, while the column density against $z$ (dashed line) is determined by integrating the SPH density from $z$ to the nearest disc surface.  

The column density estimates from the Stamatellos, Lombardi, and Combined methods are determined by multiplying the scale-height, $H$, in Figure~\ref{fig:H_estimates} by the local gas density.  It should be noted that the column density estimates from the four methods are one-sided column densities (i.e., from the $z$ to the nearest disc surface) and, hence, at the midplane are about a factor of two smaller than the full column/surface density (dotted line in Fig.~\ref{fig:column_dens_estimates}).

Comparing the column densities from the SPH simulation itself with the various estimates, shows that the Lombardi method is again unreliable near the midplane ($z \sim 0$) but produces reliable estimates at higher $z$, while the Stamatellos method produces reasonable estimates near the midplane, but produces column density estimates that are much too high at higher $z$.  Both the Combined method and the Modified Lombardi method produce reasonable estimates of the column density throughout the disc column.  

\begin{figure}
    \centering
    \includegraphics[width=0.9\linewidth]{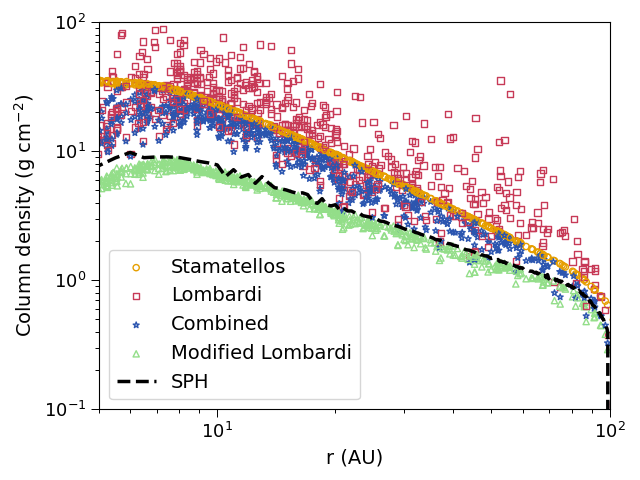}
    \caption{Column density estimates against disc radius for the $0.01$ M$_\odot$ (low-mass) disc. For the Stamatellos (yellow open circles), Lombardi (red open squares), Combined (blue stars) and Modified Lombardi (open green triangles) methods, we plot a random sample of estimates that are within $0.05$~AU of the midplane ($z = 0$). }
    \label{fig:midplane-coldens}
\end{figure}

Finally, Fig.~\ref{fig:midplane-coldens} shows the midplane column density estimates $\bar{\Sigma}(r)$ and half the disc surface density determined directly from the SPH simulation, which we take as a reasonable estimate of the column density from the disc midplane to the surface.  For the Lombardi, Stamatellos, Combined and Modified Lombardi method estimates, we plot a random sample of column density estimates that are within $0.05$~AU of the midplane ($z = 0$). 
Again, the Lombardi method shows a lot of scatter, while the combined method removes most of this scatter and produces reasonable estimates for the column density at almost all radii. However, it is clear that the Stamatellos, Lombardi, and Combined methods all slightly over-estimate the midplane column densities, with the discrepancy becoming large at small disc radii.  The Modified Lombardi method matches the expected column density at all radii, illustrating how this is a particularly reliable method for disc-like geometries.  

\subsubsection{High-mass disc}

\begin{figure}
    \centering
    \includegraphics[trim= 3.5cm 0cm 3.5cm 0cm, clip, width=0.49\linewidth]{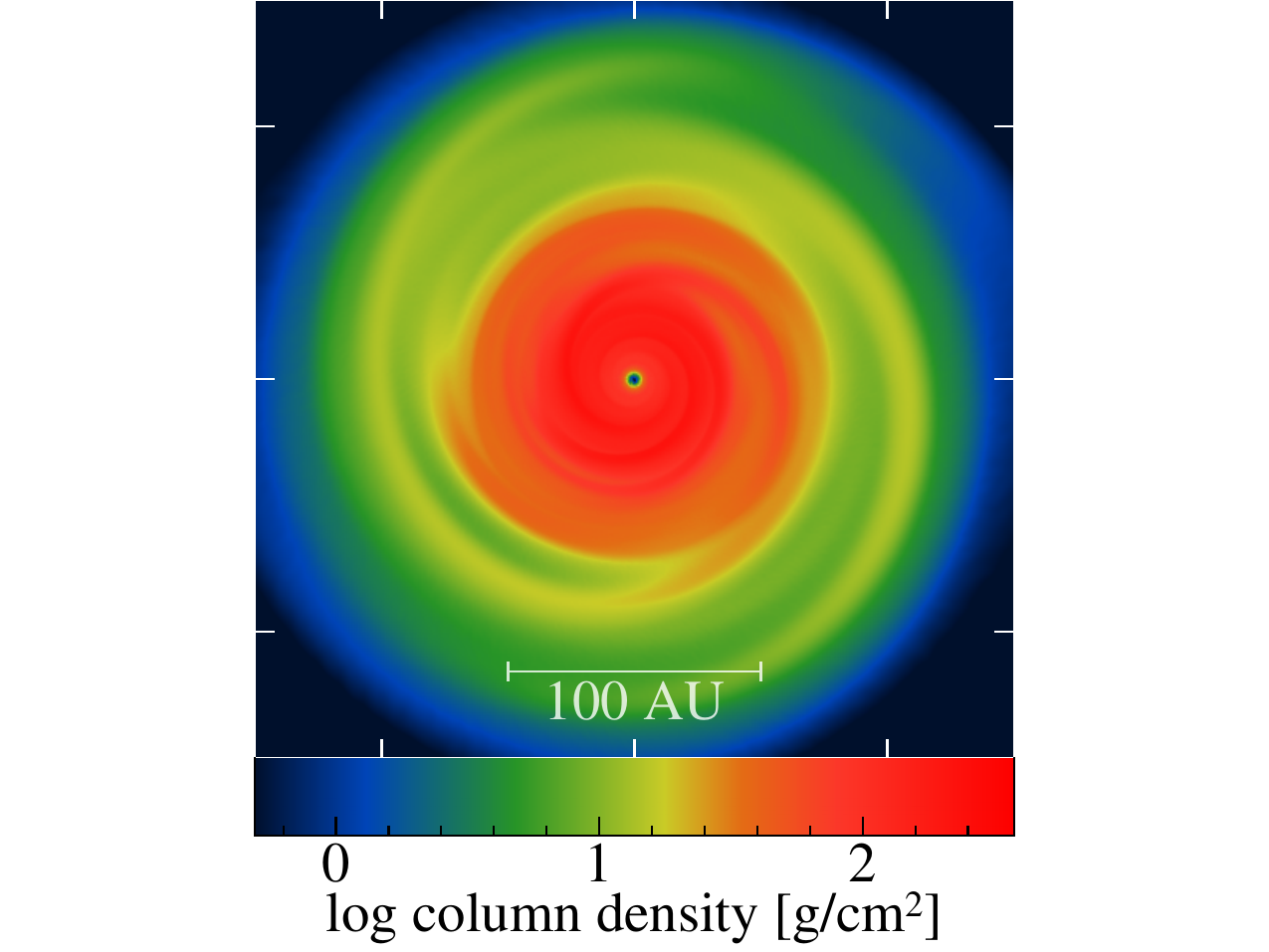}
    \hfill
    \includegraphics[trim= 3.5cm 0cm 3.5cm 0cm, clip,width=0.49\linewidth]{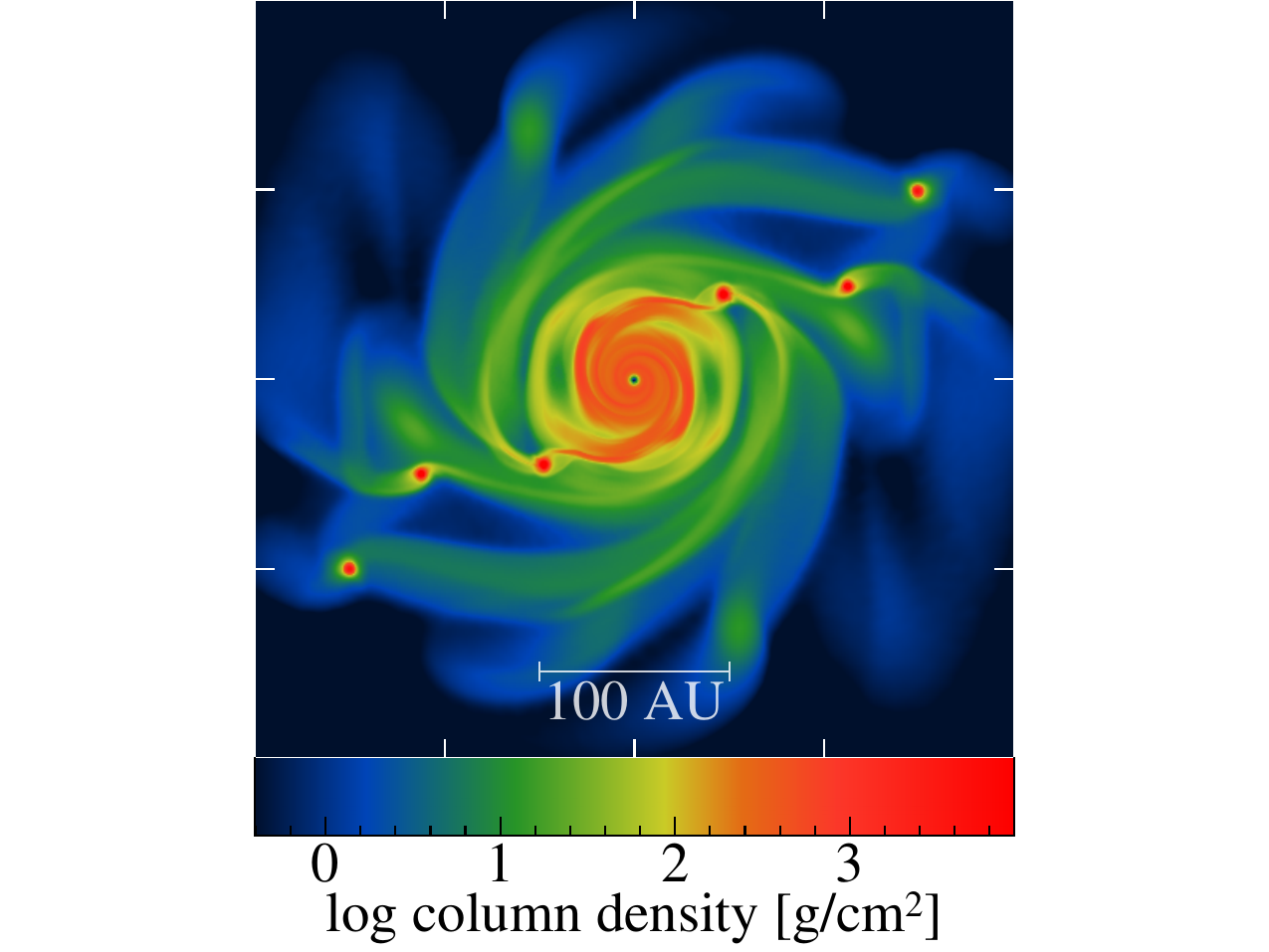}
    \caption{Column density renderings of the SPH simulations of high-mass discs with masses of $M_{\rm disc} = 0.2$~M$_\odot$ (left-hand panel) and $M_{\rm disc} = 0.25$~M$_\odot$ (right-hand panel). Both were evolved using the Combined method.}
    \label{fig:high-mass-sgdisc}
\end{figure}

We now consider two different high-mass discs, both evolved using the Combined method.  Both discs initially extend from $r = 5$~AU to $r = 100$~AU, have surface density profiles of $\Sigma \propto r^{-1}$, sound speed profiles of $c_{\rm s} \propto r^{-1/4}$, initial scale-heights at $r = 5$~AU of $H = 0.275$~AU (i.e., $H/r = 0.055$), and have central star masses of $M_* = 0.8$~M$_\odot$.  The disc masses are $M_{\rm disc} = 0.2$~M$_\odot$ and $M_{\rm disc} = 0.25$~M$_\odot$, and each disc is represented by $2 \times 10^6$ SPH particles. The artificial viscosity parameters are $\alpha_{\rm SPH}$ = 1.0 and $\beta_{\rm SPH} = 2.0 $ and a floor temperature of $T_0=10$~K was imposed. Fig.~\ref{fig:high-mass-sgdisc} shows that the $M_{\rm disc} = 0.2$~M$_\odot$ settles into a quasi-steady state with spirals, while the $M_{\rm disc} = 0.25$~M$_\odot$ disc becomes violently unstable and undergoes fragmentation.     

\begin{figure}
    \centering
    \includegraphics[width=0.9\linewidth]{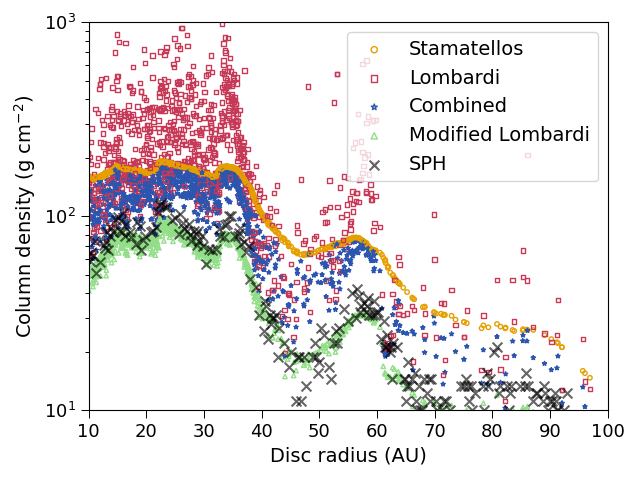}
    \caption{Column density against radius along a thin strip on the $x$-axis of the $M_{\rm disc} = 0.2$~M$_\odot$ disc shown in the left-hand panel of Figure~\ref{fig:high-mass-sgdisc}, close to the midplane. The black crosses show the column density determined by integrating the gas density in the SPH simulation, while the other symbols are column densities from the Stamatellos method (yellow open circles), the Lombardi method (red open squares), Combined method (blue stars), and Modified Lombardi method (open green triangles).}
    \label{fig:M02_coldens_vs_R}
\end{figure}

We now consider how well the different methods for estimating the column density represent structures within the disc.  Fig.~\ref{fig:M02_coldens_vs_R} shows the midplane column density along a thin strip along the positive $x$-axis of the $M_{\rm disc} = 0.2$~M$_\odot$ disc shown in the left-hand panel of Fig.~\ref{fig:high-mass-sgdisc}. The region considered is $x > 0$ AU, $|y| < 1$ AU.  The black crosses show the ``true'' column density determined by integrating the gas densities in the SPH simulation, while the symbols are the estimates from the Lombardi, Stamatellos, Combined and Modified Lombardi methods for $|z| < 0.1$ AU.   

Since the midplane is a region where the pressure gradient will tend to zero, the Lombardi method again shows a lot of scatter. The Combined method (blue stars) removes most of this scatter and seems to do reasonably well at capturing the spiral structure in the disc, for example it captures the trough at $r \sim 45$~AU and the peak at $r \sim 55$~AU.  However, as in Fig.~\ref{fig:midplane-coldens}, the Stamatellos, Lombardi and Combined methods tend to over-estimate the midplane column density, generally by a factor of about 2.  The Modified Lombardi method not only captures the midplane density structures, but also mostly matches the expected column density. 

\begin{figure}
    \centering
    \includegraphics[width=0.9\linewidth]{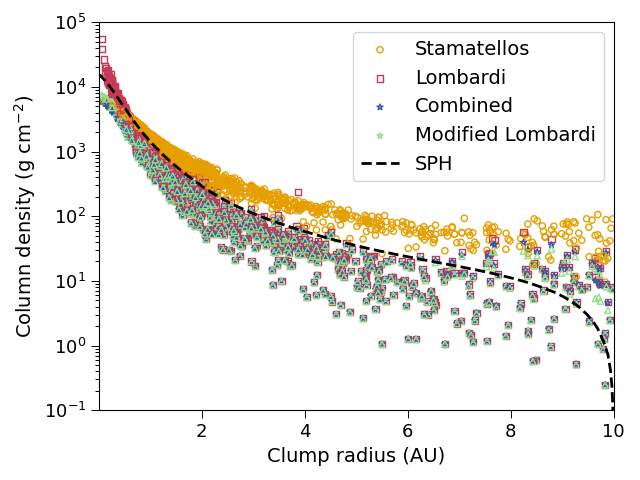}.
    \caption{Column density of one of the clumps in the fragmenting high-mass disc shown in the right-hand panel of Figure~\ref{fig:high-mass-sgdisc}. The dashed line is the SPH density integrated from the clump centre to $r_{\rm clump} = 10$~AU. The symbols are, again, a randomly selected sample of the column density estimates using the Stamatellos method (yellow open circles), Lombardi method (red open squares), Combined method (blue stars), and the Modified Lombardi method (open green triangles). The Combined and Modified Lombardi methods give very similar values near the centre of the clump and they agree with the Stamatellos method.}
    \label{fig:clump-coldens}
\end{figure}

Turning now to fragmenting discs, we examine how well the methods can recover the structure of clumps within a disc. The clump considered here formed in the $M_{\rm disc} = 0.25$~M$_\odot$ disc (Fig.~\ref{fig:high-mass-sgdisc}, right panel) and is located at $x \sim 45$~AU, $y \sim 45$~AU in the snapshot. The column density as a function of the radius from the centre of this clump is presented in Fig.~\ref{fig:clump-coldens}. Since we don't know the exact boundary of the clump, we calculate the ``true '' column density (black dashed line) by integrating the gas density out to a clump radius of $10$~AU, which could slightly over-estimate the column density as it may include some of the surrounding gas disc. The Lombardi method over-estimates the column density near the clump centre, while at larger radii, the Stamatellos methods produces column density estimates that are too high. Combining the two methods seems to produce estimates that are reasonably close to that expected. In this case, the Modified Lombardi method produces estimates very similar to those from the Combined method.  

\begin{figure}
    \centering
    \includegraphics[width=0.9\linewidth]{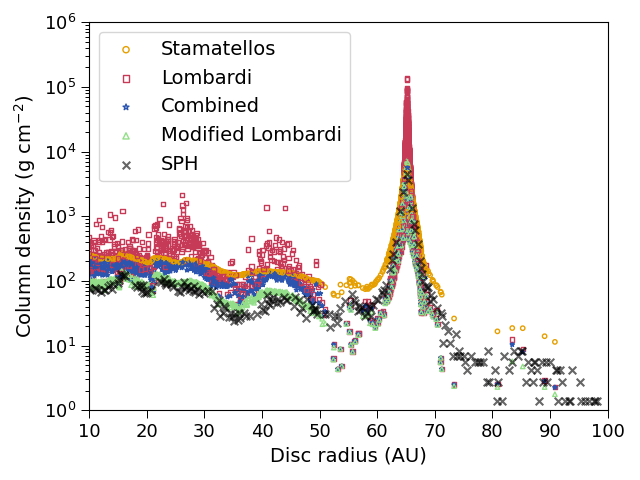}
    \caption{Cut through the disc along a radial line that includes the clump shown in Figure \ref{fig:clump-coldens} (the $M_{\rm disc} = 0.25$~M$_\odot$) showing the column densities from the SPH simulation (black crosses) and the estimate from the Stamatellos method (yellow open circles), the Lombardi method (red open squares), the Combined method (blue stars), and the Modified Lombardi method (open green triangles). For $r>50$~au, the points for the Modified Lombardi method mostly lie on top of those of the Combined method.}
    \label{fig:clump-cut-through}
\end{figure}

Finally, Figure \ref{fig:clump-cut-through} shows a radial cut through the $M_{\rm disc} = 0.25$~M$_\odot$ disc (Fig.~\ref{fig:high-mass-sgdisc} right panel) that includes the clump at $r \sim 65$~AU, the column density of which was shown in Fig.~\ref{fig:clump-coldens}. All the methods capture the clump but, again, the Lombardi method over-estimates the column density near the clump centre, while the Stamatellos method slightly over-estimates the column density in the wings of the clump.  The Combined and Modified Lombardi methods both closely recover the peak of the clump and the clump profile.

\section{Summary}

We have described and tested two new methods for approximating radiative cooling in hydrodynamical models in discs. The Stamatellos method, estimating the optical depth via the gravitational potential, has been shown to overestimate the column density, and therefore underestimate the cooling rate in discs \citep{mercer18}. The Lombardi method, in which the column density derives from the local quantity of the pressure gradient, provides a more accurate estimate of the radiative cooling rate \citep{lombardi2015} but we show that the column density estimates become unreliable close to the disc midplane. These issues are resolved by combining the two methods, to achieve improved estimates in regions of both high and low optical depth. This method is particularly suited to spherical geometries. For disc geometries, we present the Modified Lombardi method, in which the Lombardi estimate is adjusted close to the disc midplane using an estimate of the scale height in a self-gravitating disc. Whereas \citet{mercer18} show that the Lombardi and Stamatellos methods both fail to provide accurate estimates of the optical depth in a disc with clumps, the Modified Lombardi method closely recovers the column density throughout a clumpy disc and traces the peaks and troughs associated with clumps and spirals well. Of course, since the optical depth is estimated from local disc properties, more complex geometries will not be modelled well. Full radiative transfer is required to include the effects of shadowing, for example.

It is straightforward to include the effects of a central star or stars, external stellar companions and/or a variable interstellar radiation field since the background temperature can incorporate arbitrary external heating. Radiative transfer can be incorporated via the flux limited diffusion approximation as formulated by \citet{forgan2009}, since the implementation of the new methods is identical to that of \citet{stamatellos07}. {\rev A future development will be to consider the heating and gravitational influence of multiple sink particles.}

The Modified Lombardi method is ideal for modelling self-gravitating discs and presents exciting prospects for studying young protoplanetary discs with a physical treatment of cooling.

\section*{Acknowledgements}
We thank the anonymous reviewer for their kind comments that helped improve the manuscript. This research made use of the DiRAC Data Intensive service at Leicester, operated by the University of Leicester IT Services, which forms part of the Science and Technology Facilities Council (STFC) DiRAC HPC Facility (\url{www.dirac.ac.uk}). The equipment was funded by BEIS capital funding via STFC capital grants ST/K000373/1 and ST/R002363/1 and STFC DiRAC Operations grant ST/R001014/1. DiRAC is part of the National e-Infrastructure. AKY and KR are grateful for support from the UK STFC via grant ST/V000594/1. MC was supported by the Gates Cambridge Scholarship. RAB thanks the Royal Society for their support via a University Research Fellowship. EC acknowledges support from STFC grant ST/X508329/1 and DS from STFC grant ST/Y002741/1. This work made use of {\sc numpy} \citep{harris2020}, {\sc matplotlib} \citep{hunter2007} and {\sc splash} \citep{price2007}. Fig.~\ref{fig:cloud_collapse} made use of WebPlotDigitizer \citep{Rohatgi2022}. 

\section*{Data Availability}
The code developed for this paper and used to perform the simulations is available from \url{https://github.com/alisonkyoung1/phantom}, which was forked from {\sc phantom} (\url{https://github.com/danieljprice/phantom}). The simulation setup files are available on request.



\bibliographystyle{mnras}
\bibliography{references} 




\bsp	
\label{lastpage}
\end{document}